\documentclass{article}

\usepackage{natbib}
\usepackage[final]{nips_2016}

\usepackage[utf8]{inputenc} 
\usepackage[T1]{fontenc}    
\usepackage{hyperref}       
\usepackage{url}            
\usepackage{booktabs}       
\usepackage{amsfonts}       
\usepackage{nicefrac}       
\usepackage{microtype}      

\usepackage{amsmath}
\usepackage{amsthm}
\usepackage{bbm}
\usepackage{mathabx}

\usepackage{subfigure}
\usepackage{graphicx}

\DeclareMathOperator*{\argmax}{argmax}
\DeclareMathOperator*{\argmin}{argmin}
\DeclareMathOperator{\sign}{sign}

\title{Efficiency of active learning for the allocation of workers on crowdsourced classification tasks}

\author{
  Edoardo Manino\\
  University of Southampton\\
  \texttt{em4e15@soton.ac.uk}\\
  \And
  Long Tran-Thanh\\
  University of Southampton\\
  \texttt{ltt08r@ecs.soton.ac.uk}\\
  \And
  Nicholas R. Jennings\\
  Imperial College, London\\
  \texttt{n.jennings@imperial.ac.uk}\\
}

\begin{document}

\maketitle

\begin{abstract}
Crowdsourcing has been successfully employed in the past as an effective and cheap way to execute classification tasks and has therefore attracted the attention of the research community. However, we still lack a theoretical understanding of how to collect the labels from the crowd in an optimal way. In this paper we focus on the problem of worker allocation and compare two active learning policies proposed in the empirical literature with a uniform allocation of the available budget. To this end we make a thorough mathematical analysis of the problem and derive a new bound on the performance of the system. Furthermore we run extensive simulations in a more realistic scenario and show that our theoretical results hold in practice.
\end{abstract}

\section{Introduction}

Crowdsourcing has emerged as an effective way of hiring a temporary workforce from an online community and execute a large number of simple and repetitive data-processing tasks. Despite the unreliability of the single members, as a whole these crowds have been proven capable of producing valuable solutions that would not have been achievable with traditional methods. Our focus in particular is on classification projects, where the employer presents the workers with a series of multiple-choice questions and use their answers to infer the correct label of the tasks. This model has been successfully employed for various purposes including image labeling \citep{Sorokin2008}, ballot voting in complex workflows and filtering \citep{Parameswaran2012}.

As the market and the demand for these kind of solutions expand, there is an increasing interest in improving the efficiency of the crowdsourcing process. In the case where the workers are rewarded with small payments, as in the famous Amazon Mechanical Turk\footnote{\texttt{mturk.com}} platform, this goal is particularly important as it could make crowdsourcing even more attractive as a cheap alternative. One of the main obstacles however is the heterogeneity in the quality of the answers collected from the crowd \citep{Ipeirotis2010}. As a large number of them may be incorrect, the usual strategy consists in letting several workers execute the same task and then average their answers to come up with a reliable conclusion. For this reason much of the existing research in machine learning techniques for crowdsourcing has thus focussed on how to optimally aggregate the set of labels collected from the crowd and reduce the number of misclassified tasks.

Our research adresses a related problem. During the crowdsourcing process itself in fact, we need to allocate the available workforce on the individual tasks. However, the decisions we take are likely to influence the final performance of the aggregation method in place. The research question is thus finding which allocation policies guarantee the best results at the end of the crowdsourcing effort. Perhaps surprisingly, this problem has received less attention in the existing literature and while there are some empirical results, little is known about their theoretical properties \citep{Slivkins2013}. In this paper we investigate those properties, though in a simplified setting, and show that they hold even in a more realistic scenario.

In next section we introduce our crowdsourcing model and the related works of the existing literature. Then we present our theoretical analysis of active learning policies in an ideal scenario where the quality of each individual worker is known in advance. Finally we show that the empirical performance of active learning does not degrade when this piece of information is not accessible and an inference algorithms is used to estimate it.

\section{Background} 

We analyse here a setup where we have to discover the true label $\ell^*_i$ of a set of independent tasks $i\in[1,M]$. These labels are binary and we conventionally restrict their values in the set $L=\{+1,-1\}$. We further assume that we have a maximum budget $B$ of votes that we can gather from the crowd, and the individual vote $\ell_{ij}$ of worker $j$ on task $i$ has probability $\mathbbm{P}(\ell_{ij}=\ell^*_i)=p_j$ of being correct.

In other terms we model the accuracy of each worker as a single parameter that is neither affected by the true label of the task she is working on, nor by other factors such fatigue effect or tasks with different degree of difficulty. In this model, \emph{experts} will have a $p_j$ close to one, whereas \emph{spammers} are associated to a $p_j\approx1/2$. On the other side of the spectrum, we call \emph{adversaries} the workers with $p_j<1.2$ as their answers are likely to be incorrect.

When the parameters $p_j$ of the workers are known, we can optimally aggregate the workers' labels according to the following rule \citep{Nitzan1982}:

\begin{equation}
\hat{\ell}_i=\sign\bigg\{\sum_{j\in\mathcal{N}_i}\ell_{ij}\log\Big(\frac{p_j}{1-p_j}\Big)\bigg\}
\label{eq:nitzan}
\end{equation}
where $\mathcal{N}_i$ is the subset of workers who executed task $i$. Notably, these rule follows directly from the log-ratio of the likelihood of each class and allows us to compute the confidence in our prediction as $\mathbbm{P}(\hat{\ell}_i=\ell^*_i)=\exp(|z_i|)/(1+\exp(|z_i|))$ where $z_i$ is the argument of the sign function in Equation \ref{eq:nitzan}.

In a more realistic scenario though, the level of expertise of each individual worker is unknown and thus we need to estimate it. A common idea is to use some machine learning algorithm to infer both the parameters $p_j$ and the task classification at the same time in an unsupervised fashion. The first application of this approach dates back to the expectation-maximisation algorithm of \citet{Dawid1979} but in more recent years many alternative techniques have been proposed \citep{Karger2011,Zhou2012}. In this paper we will use the approximate mean-field algorithm proposed by \citet{Liu2012} which consists in iterating between the following two steps until convergence:

\begin{align}
\texttt{E-step: } & \mu_i(\ell)\propto\prod_{j\in\mathcal{N}_i}\hat{p}_j^{\delta_{ij}}(1-\hat{p}_j)^{1-\delta_{ij}} \\
\texttt{M-step: } & \hat{p}_j=\frac{\sum_{i\in\mathcal{N}_j}\mu_i(\ell_{ij})+\alpha}{|\mathcal{N}_j|+\alpha+\beta}
\end{align}
where $\delta_{ij}=1$ if $\ell_{ij}=\ell$ and zero otherwise, while $\alpha$ and $\beta$ are the parameters of the Beta distribution which is used as the prior on the workers' skill.

These inference techniques are designed to minimise the number of misclassification errors given a set of labels, but they give no indication on how to conduct the label collection process itself. The only exception is the work by \citet{Karger2011}, who propose to assign the same number $r=B/M$ of workers to each task and prove that this policy is order-optimal with respect to any other allocation policy that uses only the information available before the crowdsourcing process.

In contrast, a number of empirical-oriented works on crowdsourcing advocate the use of other policies that adapt the allocation depending on the data collected during the crowdsourcing process. Specifically \citet{Welinder2010} propose an algorithm that computes the confidence in the classifications after the acquisition of every new label, and allocates more workers on the most uncertain tasks. \citet{Simpson2014} address the same problem from a different perspective and introduce a greedy algorithm to maximise the amount of information collected from the crowd. Notably these two approaches belong to the family of \emph{active learning} policies \citep{Settles2010}, but the authors provide no theoretical guarantee on their performances.

\section{A theoretical analysis of allocation policies}

In this section we analyse the impact that different worker allocation strategies have on the performance of the system. Specifically we are interested in computing the probability $\mathbbm{P}(\hat(\ell)_i=\ell^*_i)$ of classifying task $i$ correctly under each of the following policies:

\begin{itemize}
\item the \emph{uniform allocation} policy which allocates the same number of workers to each task \citep{Karger2011};
\item a greedy policy that maximises the expected information gain of each new incoming label \citep{Simpson2014};
\item a simple \emph{uncertainty sampling} rule that assigns new workers on the task with the least confident label \citep{Lewis1994, Welinder2010}.
\end{itemize}

To keep our discussion simple, we model our interaction with the crowd as a sequence of $B$ distinct rounds. In each round a new worker $j$ becomes available and we have to assign her to a task $i$ of our choice. Once the label $\ell_{ij}$ is acquired, the system moves on to the next round $j+1$. Furthermore we assume the presence of an \emph{oracle} that at each round reveals the parameter $p_j$ of the incoming worker. We will relax this assumption in Section \ref{sec:inference}.

\subsection{Equivalence between different active learning policies}

We can now adapt the worker allocation policy of \citet{Simpson2014} to our setup by choosing the task $i$ that maximises the amount of information collected at each round. To do so we interpret the effect of the next incoming label as a random variable $x_j=\pm \log(p_j/(1+p_j))$ which modifies the current posterior probability $\mathbbm{P}(\ell^*_i=+1)=\exp(z_i)/(\exp(z_i)+1)$, and measure its information gain as the Kullback-Leibler divergence between the two distributions:

\begin{equation}
\label{eq:inf_gain}
\mathcal{I}(i,x_j)=\frac{x_j\exp(z_i+x_j)}{\exp(z_i+x_j)+1}+\log\left(\frac{\exp(z_i)+1}{\exp(z_i+x_j)+1}\right)
\end{equation}

As the sign of $x_j$ is unknown a priori, we need to choose the task $i$ that maximises Equation \ref{eq:inf_gain} in expectation, i.e. $i^*=\argmax_i(\mathbbm{E}_{x_j}(\mathcal{I}(i,x_j)))$. Fortunately it is possible to perform such computation explicitly by taking into account the current posterior on $i$ and the probability $p_j$ of observing the correct label. Moreover it can be shown that the expected information gain is a symmetric function around its maximum in $z_i=0$ for any $\log(p_j/(1-p_j))\neq0$ (we omit the derivation here for lack of space). As a consequence, we can make the optimal greedy decision using the following rule:

\begin{equation}
i^*=\argmin_i(|z_i|)
\label{eq:magic}
\end{equation}

Interestingly, Equation \ref{eq:magic} corresponds to the uncertainty sampling rule, as $i^*$ is the task with the least confident label. We have therefore proven that the two active learning policies are actually the same, and as a result we will consider them as a single one in the following discussion.

\subsection{A random walk interpretation of the allocation policies}

From the point of view of a single task $i$, the active learning policy keeps allocating new workers in short bursts of activity, until $i$ is no more the least confident. In this respect, the evolution of the log-likelihood ratio $z_i$ can be interpreted as a one-dimensional random walk with starting point $z_i=0$. Furthermore, at the end of the crowdsourcing process we can identify a confidence threshold $z_B\in\mathbbm{R}$ such that $z_B=\max_x\{|z_i|\geq x\}$ for all the task $i\in[1,M]$. The random walk has thus the property of being \emph{bounded} in that whenever $|z_i|$ reaches $z_B$, task $i$ does not receive any additional worker.

\begin{figure}[tb]
  \centering
  \subfigure[]{
    \includegraphics[width=0.45\textwidth]{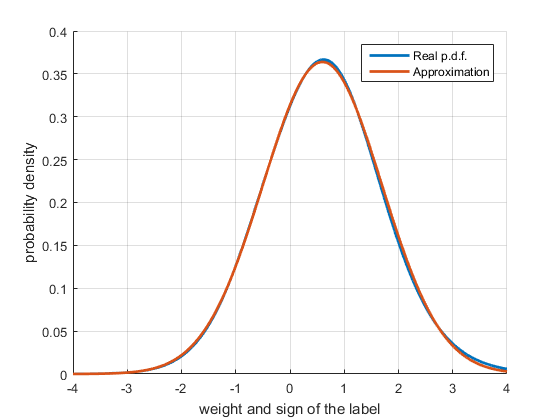}
    \label{fig:population_1}
  }
  \subfigure[]{
    \includegraphics[width=0.45\textwidth]{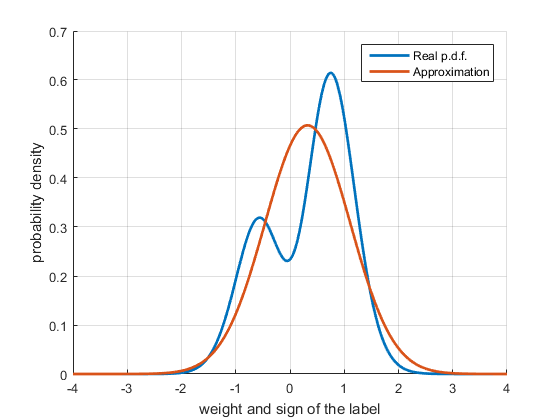}
    \label{fig:population_2}
  }
  \caption{Probability density function in the log-likelihood ratio domain and its normal approximation, (a) a broad-shaped Beta(4,2) and (b) a Beta(16,8) where the two modes become visible.}
  \label{fig:population}
\end{figure}

Every new label collected from the crowd corresponds to step in the random walk. Given the probability distribution $f_P(p)=\mathbbm{P}(p_j=p)$ of the workers' skill, we can derive the distribution of the weights $w_j=\log(p_j/(1-p_j))$ associated to the vote of each worker as follows:

\begin{equation}
f_W(z)=\frac{\exp(z)}{(1+\exp(z))^2}f_P\Big(\frac{\exp(z)}{1+\exp(z)}\Big)
\end{equation}
and then compute the p.d.f. of the labels $x_j=\pm w_j$ received from the crowd by assuming that $\ell^*_i=+1$ and taking into account that $\mathbbm{P}(\ell_{ij}=\ell^*_i)=\exp(w_j)/(1+\exp(w_j))$:

\begin{equation}
f_V(z)=\frac{\exp(z)}{1+\exp(z)}(f_W(z)+f_W(-z))
\end{equation}
an example of which can be found in Figure \ref{fig:population}.

With a closed form or a numerical representation of $f_V$ we can iteratively compute the expected number of steps the random walk needs to reach one of the two boundaries $\pm z_B$. Specifically we have to perform the following computation:

\begin{align}
f_A^1(z) &= f_V(z)\\
f_B^k(z) &= f_A^k(z)\mathbb{I}(z\in(-z_B,z_B))\\
f_A^{k+1}(z) &= (f_B^k\ast f_V)\\
\mathbbm{E}(r_a) &= \sum_{k=1}^\infty k\Big(\int_{-\infty}^{-z_B} f_A^k(z)dz+\int_{+z_B}^{\infty} f_A^k(z)dz\Big)
\label{eq:act_step}
\end{align}
where $\mathbbm{I}(\bullet)$ is the indicator function and $\ast$ the convolution operator.

In contrast, we can interpret the uniform allocation case as an \emph{unbounded} random walk where we fix the number of steps $r_u$ and compute the probability of a correct classification as follows:

\begin{equation}
\mathbbm{P}(\hat{\ell}_i^u=\ell^*_i)=\int_{0^+}^\infty (\Asterisk_{r_u} f_v(z))dz
\label{eq:uni_conf}
\end{equation}

A comparison between Equation \ref{eq:act_step} and \ref{eq:uni_conf} is possible by finding a value of $z_B$ such that $\mathbbm{E}(r_a)=r_u$. As can be seen in Figure \ref{fig:conv_comp} for a population $f_P\sim\text{Beta}(4,2)$, the active learning policy provides greater accuracy except for small values of $r_u$.

\subsubsection{The homogeneous crowd case}

An interesting scenario arises when all the workers have the same parameter $p_j=p$, i.e. the distribution of the crowd is a Dirac delta $f_P=\delta(p)$. In this case the labels collected from the crowd have the same weight and thus our aggregation technique simplifies to a \emph{majority voting} rule. Moreover we can compute a closed form of both Equations \ref{eq:act_step} and \ref{eq:uni_conf}, with the only assumption that $r_u$ is an odd integer:

\begin{align}
\mathbbm{E}(r_a)&=\frac{\big(2\frac{\exp(z_B)}{1+\exp(z_B)}-1\big)z_B}{(2p-1)\log\big(\frac{p}{1-p}\big)}\\
\mathbbm{P}(\hat{\ell}_i^u=\ell^*_i)&=\sum_{r=\lceil r_u/2\rceil}^{r_u} p^r(1-p)^{r_u-r}
\end{align}

\begin{figure}[tb]
  \centering
  \subfigure[]{
    \includegraphics[width=0.45\textwidth]{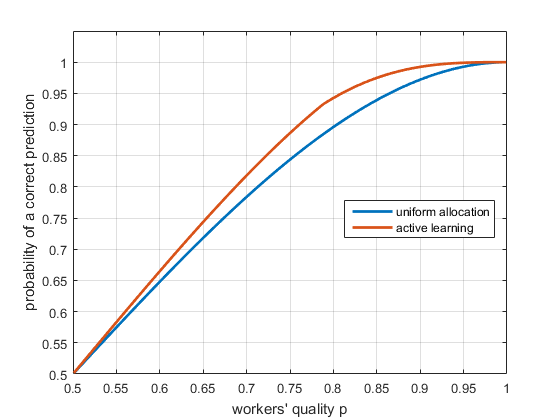}
    \label{fig:theor_comp_1}
  }
  \subfigure[]{
    \includegraphics[width=0.45\textwidth]{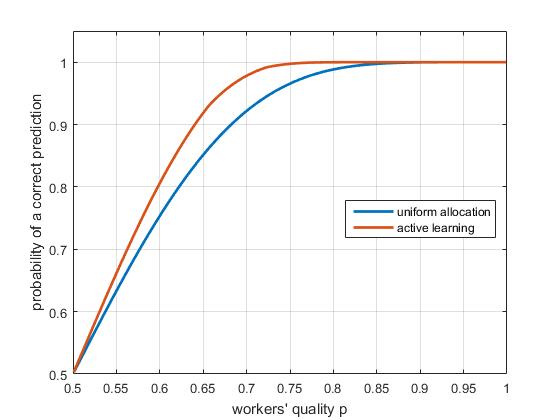}
    \label{fig:theor_comp_2}
  }
  \caption{Comparison between uniform allocation and active learning in the homogeneous case with an average of (a) $r_u=3$ votes per task and (b) $r_u=11$ votes per task.}
  \label{fig:theor_comp}
\end{figure}

Interestingly, for any value of $r_u > 1$ and $p\in(0,1/2)\cup(1/2,1)$, the active learning policy achieves better performances than uniform allocation as is shown in Figure \ref{fig:theor_comp}.

\subsubsection{Probability bounds on the heterogeneous case}

More in general, we may have only access to the first moments of the distribution $f_V$. If this is the case we can still provide some closed-form bounds on Equation \ref{eq:act_step} and \ref{eq:uni_conf}. First of all we need to assume that the distribution $f_V$ is bounded in $[-\gamma,\gamma]$ with $\gamma<\infty$ or, in other terms, that there are no perfect workers with $p_j=1$ or $p_j=0$ in our population. We can then apply the results in \citet{Ethier2002} and derive an upper bound on the expected number of labels required by the active learning policy to reach $z_B$:

\begin{equation}
\mathbbm{E}(r_a) \leq \frac{1}{\mathbbm{E}(f_V)}\Big((2z_B+\gamma)\frac{\rho_0^{z_B+\gamma}-1}{\rho_0^{2z_B+\gamma}-1}-z_B\Big)
\label{eq:gambler_ruin}
\end{equation}
where $\rho_0\neq1$ is the solution of the equation $\mathbbm{E}(\rho^{f_V})=1$ and can be approximated as follows:

\begin{equation}
\rho_0\approx\frac{2\mathbbm{E}(f_V)}{\text{Var}(f_V)}
\end{equation}
which does hold as long as $f_V$ is close to a normal distribution (see Figure \ref{fig:population} for an example) \citep{Kozec1995,Canjar2007}.

In the case of uniform allocation instead, we can observe that $z_i$ is the sum of $r_u$ identically-distributed independent variables and apply Chernoff's inequality to bound Equation \ref{eq:uni_conf}:

\begin{equation}
\mathbbm{P}(\hat{\ell}_i^u=\ell^*_i) \leq \frac{1}{1+\frac{r_u(\mathbbm{E}(f_V))^2}{\text{Var}(f_V)}}
\label{eq:concentration}
\end{equation}

\begin{figure}[tb]
\centering
\includegraphics[width=0.6\textwidth]{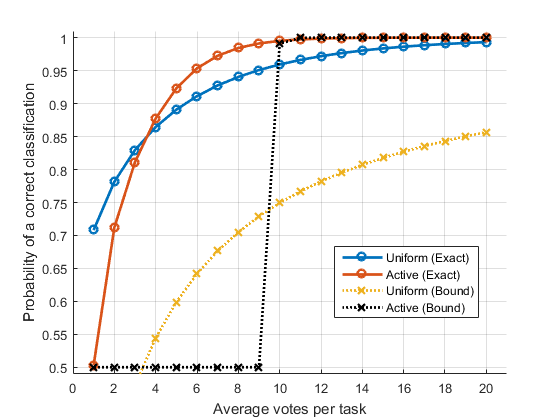}
\caption{Comparison between uniform allocation, active learning and their bounds with a Beta(4,2) population and increasing values of $r_u$.}
\label{fig:conv_comp}
\end{figure}

It must be noted that Equation \ref{eq:concentration} is not asymptotically tight, but nevertheless provides a better bound for ``small'' values of $r_u$ compared to other concentration inequalities \citep{Hoeffding1963}. A comparison between the bounds of the two policies is shown in Figure \ref{fig:conv_comp} where we assumed $\gamma=5\sqrt{\text{Var}(f_V)}$: as long as $r_u>\gamma/\mathbbm{E}(f_V)$ the bound on the active learning policy is quite tight, but some work is still necessary to improve it in the other case.

\section{An empirical comparison between policies}
\label{sec:inference}

In this section we study a more realistic scenario where the parameters $p_j$ of the workers are unknown and the workers are allowed to execute multiple tasks. These two differences in the setting have the following consequences:

\begin{itemize}
\item in order to aggregate the set of labels collected from the crowd we need to rely on the output of some inference method, which is inherently less reliable than the prediction of the optimal rule by \citet{Nitzan1982};
\item in general, having the workers execute multiple tasks is beneficial as it improves the accuracy of the estimation of their skills, however it can also interfere with the allocation policy as we cannot assign the same worker $j$ on a particular task $i$ more than once.
\end{itemize}

In order to test the performances of the different policies we had to make a couple of changes in our system with respect to the previous section. First, we selected the approximate mean-field algorithm by \citet{Liu2012} to learn the workers' parameter $p_j$ and predict the classification of the tasks. As this algorithm needs a prior on the distribution $f_P$, we decided to test the most favourable scenario where the prior matches the real distribution. Second, in order to avoid any repetition of a worker-task pair $(j,i)$ we modified the allocation policies as follows:

\begin{align}
\text{Uniform allocation: } & i^*=\min_{i:j\not\in\mathcal{N}_i}\{|\mathcal{N}_i|\}\\
\text{Active learning: } & i^*=\min_{i:j\not\in\mathcal{N}_i}\{|z_i|\} 
\end{align}
where $z_i$ is estimated by running one iteration of an online version of the approximate mean-field algorithm after the arrival of each new label.

In our experiments we explored the impact of three parameters on the performance of our crowdsourcing system: the number of tasks $M$, the total budget $B$ and the average number of labels $|\mathcal{N}_j|$ provided by each worker. All the results we provide are generated by extracting the workers' skills from a distribution $f_P\sim\text{Beta}(4,2)$ and simulating $B$ rounds of the crowdsourcing process. For each datapoint we show the estimated mean and standard error after 1000 simulations.

\begin{figure}[htb]
\centering
\includegraphics[width=0.6\textwidth]{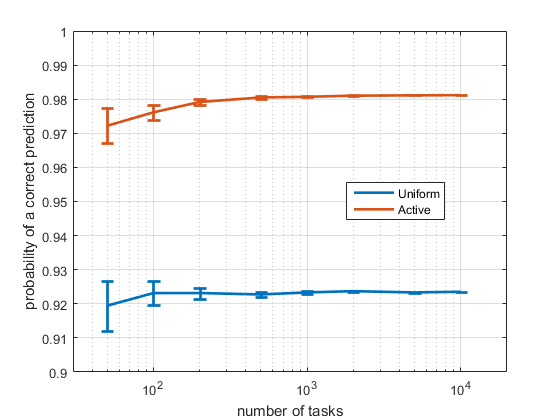}
\caption{Classification accuracy of uniform allocation and active learning on different sizes of the crowdsourcing project.}
\label{fig:tasks}
\end{figure}

As far as the number of tasks is concerned, we tested a wide range of values from $M=20$ to $M=10000$. For this set of experiments the budget was set to $B=10M$ and the number of labels per worker to $|\mathcal{N}_j|=10$. As the data in Figure \ref{fig:tasks} show, the performances of the two allocation policies are similar for all the values of $M$.

\begin{figure}[htb]
\centering
\includegraphics[width=0.6\textwidth]{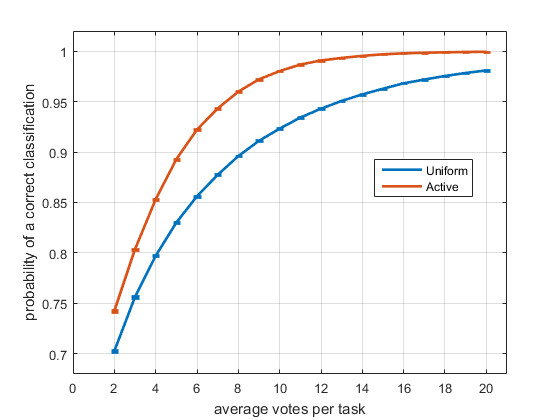}
\caption{Classification accuracy of uniform allocation and active learning with different budget $B$.}
\label{fig:budget}
\end{figure}

In contrast the amount of budget $B$ has a strong impact on the accuracy of the system as a larger number of labels improves the probability of a correct classification. Figure \ref{fig:budget} shows the results with $M=1000$, $|\mathcal{N}_j|=10$ and values of $B/M$ ranging from 2 to 20. It is interesting to notice that while the active learning policy achieves a better performance than uniform allocation, the gap between the two is greater than what the theoretical analysis of the previous section may suggests (see Figures \ref{fig:theor_comp} and \ref{fig:conv_comp}).

\begin{figure}[tb]
\centering
\includegraphics[width=0.6\textwidth]{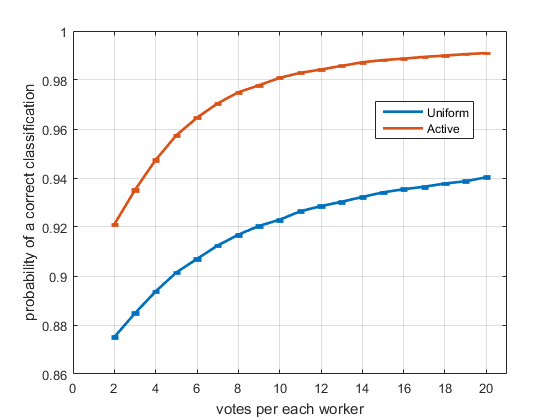}
\caption{Classification accuracy of uniform allocation and active learning with different number of labels provided by each worker.}
\label{fig:productivity}
\end{figure}

Finally, the number of labels provided by each worker has an impact on the performance of the system. Even with a fixed number of tasks $M=1000$ and budget $B=10M$, the accuracy of the predictions improves as $|\mathcal{N}_j|$ increases from 2 to 20 as shown in Figure \ref{fig:productivity}. This phenomenon is most likely due to the fact that the inference algorithm is able to exploit the larger amount of data available per each worker, and produce better estimates of the parameters $p_j$.

\section{Conclusions and future works}

In this paper we addressed the problem of worker allocation on crowdsourced classification tasks from a theoretical perspective. First we studied the properties of three allocation policies proposed by the existing literature. During our analysis we were able to prove the equivalence between two different active learning strategies and provide a new bound on their performance. Second, we compared these adaptive active learning policies with a uniform allocation one and showed that the latter is less efficient in terms of classification accuracy. Finally we validated our theoretical results through an extensive empirical analysis. Future works include improving our bounds when the available budget is small, and extending our theoretical analysis to the more realistic case where the workers' skill is unknown a priori.

\subsubsection*{Acknowledgments}

This research is funded by the UK Research Council project ``ORCHID'', grant EP/I011587/1

\end{document}